\documentstyle{article}
\begin{document}
\title{WHERE AND WHY THE GENERALIZED HAMILTON-JACOBI REPRESENTATION DESCRIBES
MICROSTATES OF THE SCHR\"{O}DINGER WAVE FUNCTION}
\author{Edward R. Floyd \\
{\it 10 Jamaica Village Road} \\
{\it Coronado, California 92118-3208}}
\date{Received March 14, 1996}
\maketitle
 
\noindent A generalized Hamilton-Jacobi representation describes microstates of
the Schr\"{o}dinger wave function for bound states.  At the very points that
boundary values are applied to the bound state Schr\"{o}dinger wave function, the
generalized Hamilton-Jacobi equation for quantum mechanics exhibits a nodal
singularity.  For initial value problems, the two representations are equivalent.
\\  


\noindent {\bf 1. INTRODUCTION} \\

The Schr\"{o}dinger equation is a linear differential equation.  Consequently,
the Schr\"{o}dinger representation of quantum mechanics is well development and
most familiar.  In the early days of quantum mechanics, the physics community had
considered that in one dimension an equivalent representation of quantum
mechanics was rendered by a generalized Hamilton-Jacobi equation [1--3] or the
related Milne [4] or Pinney [5] equation.  These equations are nonlinear
differential equations.  Consequently, their development as a representation of
quantum mechanics is not as extensive.  Yet, some workers have noted various
computational advantages for directly solving the generalized Hamilton-Jacobi
representation or its equivalent rather than solving the Schr\"{o}dinger
representation [4,6--10].  Even today familiarity with the generalized Hamilton-
Jacobi representation ends for most other workers with the WKB approximation
where the higher order terms of the generalized Hamilton-Jacobi equation have
been ignored.  There still is a widely held perception that the generalized
Hamilton-Jacobi representation is at best only equivalent to the Schr\"{o}dinger
representation and offers nothing new.   

With regard to the foundations of quantum mechanics, recent progress in the
generalized Hamilton-Jacobi representation has shown that this representation 
sometimes, albeit not always, renders microstates of the Schr\"{o}dinger wave
function.  On the one hand, the generalized Hamilton-Jacobi representation of
bound states has shown that each of various, non-unique trajectories in phase
space for energy $E$ is consistent with the unique eigenfunction of energy $E$
of the Schr\"{o}dinger representation [11].  Each of these distinct trajectories
determines a microstate of the Schr\"{o}dinger wave function [11].  These
trajectories differ with possible Feynman paths.  Each trajectory or microstate
alone determines the Schr\"{o}dinger wave function in contrast to Feynman's
giving equal weight to all possible paths whose phases are subject to a classical
generator of the motion.  These microstates undermine the widely held belief that
the Schr\"{o}dinger wave function be an exhaustive description of quantum
phenomena.  On the other hand, the phase-space trajectory for tunneling or
reflection and transmission problems does not manifest any microstates for the
trajectory is unique and consistent with the unique Schr\"{o}dinger wave function
[12,13].     
 
Herein, we resolve why a generalized Hamilton-Jacobi representation only
sometimes describes microstates for the Schr\"{o}dinger wave function of
nonrelativistic quantum mechanics.  The reasons are mathematical.  We investigate
the Schr\"{o}dinger and generalized Hamilton-Jacobi representations in one
dimension and make limited extensions to higher dimensions.  We consider the
time-independent case for both initial condition and boundary value problems. 
As a byproduct, the Schr\"{o}dinger equation and the generalized Hamilton-Jacobi
equations are shown not to be equivalent of each other.  In this investigation,
we examine the consequences of using a generalized Hamilton-Jacobi equation
rather than the Milne or Pinney equation to describe quantum phenomena.  This
choice was arbitrary as we would have produced the same findings by using the
Milne or Pinney equation. \\[.05in]

\noindent {\bf 2. THEORY} \\

Before we examine specific cases, we present established relationships between
the solutions for generalized Hamilton-Jacobi equation and the Schr\"{o}dinger
equation.  The generalized Hamilton-Jacobi equation for quantum mechanics is
given in one dimension $x$ by [3]

\begin{equation}
\frac{(W')^2}{2m}+V-E=-\frac{\hbar ^2}{4m}\left[\frac{W'''}{W'}-
\frac{3}{2}\left(\frac{W''}{W'}\right)^2 \right],
\end{equation}

\noindent where $W$ is Hamilton's characteristic function, $W'$ is the momentum
conjugate to $x, V$ is the potential, $E$ is energy, $m$ is the mass of the
particle, and $\hbar=h/(2\pi)$ where in turn $h$ is Planck's constant.  The left
side of Eq.\ (1) manifests the classical Hamilton-Jacobi equation which renders
the generator of motion for Feynman paths.  The right side of Eq.\ (1) is the
Schwarzian derivative which manifests the quantum effects.

We explicitly note that $W$ and $W'$ are real even in the classically forbidden
region.  The general solution for $W'$ is given by [11]

\begin{equation}
W'=\pm (2m)^{1/2}(a\phi ^2+b\theta ^2+c\phi \theta )^{-1},
\end{equation}

\noindent where $(a,b,c)$ is a set of real coefficients such that  $a,b > 0$, and 
$(\phi,\theta)$ is a set of normalized independent solutions of the associated
time-independent Schr\"{o}dinger equation, $-\hbar ^2\psi''/(2m) + (V-E)\psi =
0$.  The independent solutions $(\phi,\theta)$ are normalized so that their
Wronskian, ${\cal W}(\phi,\theta) = \phi \theta ' - \phi '\theta $, is scaled to
give  ${\cal W}^2(\phi,\theta) = 2m/[\hbar ^2(ab-c^2/4)] > 0$.  This ensures that
$(a\phi ^2 + b\theta ^2 + c\phi \theta) > 0$.  We note for completeness that a
particular set $(\phi,\theta)$ of independent solutions of the Schr\"{o}dinger
equation may be chosen by the superposition principle such that the coefficient
$c$ is zero.  The $\pm$ sign in Eq.\ (2) designates that the motion may be in
either $x$-direction.  Hereon, we shall assume motion in the $+x$-direction.  The
corresponding solution for the generalized Hamilton's characteristic function,
$W$, which is also a generator of the motion, is given by

\[
W=\hbar \arctan \left(\frac{b(\theta /\phi ) + c/2}{(ab-c^2/4)^{1/2}}\right)+K,
\]

\noindent where $K$ is an integration constant which we may set to zero herein.

We now show that $(\phi ,\theta)$ can only be a set of independent solutions of
the Schr\"{o}dinger equation.  Direct substitution of Eq.\ (2) for $W'$ into Eq.\
(1) renders

\begin{eqnarray}
&  & \frac{a\phi +c\theta /2}{a\phi ^2+b\theta ^2+c\phi \theta }[-\hbar ^2/(2m)
\phi ''-(E-V)\phi ] \nonumber \\
&  & + \frac{b\theta +c\phi /2}{a\phi ^2+b\theta ^2+c\phi \theta }[-\hbar ^2/(2m)
\theta ''-(E-V)\theta ] \nonumber \\
&  & - \frac{[{\cal W}^2\hbar ^2(ab-c^2/4)/(2m)-1]}{(a\phi ^2+b\theta ^2+c\phi
\theta )^2} \  = \  0.
\end{eqnarray}

\noindent For the general solution for $W'$, the real coefficients $(a,b,c)$ are
arbitrary within the limitations that  $a,b > 0$  and from the Wronskian that 
$ab - c^2/4 > 0$.  Hence, for generality the expressions within each of the three
square brackets on the left side of Eq.\ (3) must vanish identically.  The
expressions within the first two of these square brackets manifest the
Schr\"{o}dinger equation, so the expressions within these two square brackets are
identically zero if and only if $\phi$ and $\theta$ are solutions of the
Schr\"{o}dinger equation.  The expression within third bracket vanishes
identically if and only if the normalization of the Wronskian is such that 
${\cal W}^2(\phi,\theta) = 2m/[\hbar ^2(ab-c^2/4)]$.  For  ${\cal W}(\phi,\theta)
\neq 0$,  $\phi$ and $\theta$ must be independent solutions of the
Schr\"{o}dinger equation.  Hence, $\phi$ and $\theta$ must form a set of
independent solutions of the Schr\"{o}dinger equation.

Let us investigate, first, the initial condition problem.  The Schr\"{o}dinger
wave function is complex and is uniquely specified by initial conditions $\psi
(x_o)$ and $\psi '(x_o)$.  By the superposition principle, the Schr\"{o}dinger
wave function is described by  $\psi  =  \alpha \phi + \beta \theta$  where
$\alpha$ and $\beta$ are coefficients uniquely specified, as well known, by the
initial conditions as

\[
\alpha = \frac{\psi (x_o)\theta '(x_o)-\psi '(x_o)\theta (x_o)}{\phi (x_o)\theta
'(x_o)-\phi '(x_o)\theta (x_o)}
\]

and

\[
\beta = -\frac{\psi (x_o)\phi '(x_o)-\psi '(x_o)\phi (x_o)}{\phi (x_o)\theta
'(x_o)-\phi '(x_o)\theta (x_o)}.
\]

\noindent The coefficients $\alpha$ and $\beta$ may be complex.  The steady-state
Schr\"{o}dinger wave function can also be expressed by [12]

\begin{eqnarray}
\psi & = & \frac{(2m)^{1/4}}{(W')^{1/2}[a-c^2/(4b)]^{1/2}} \exp (iW/\hbar )
\nonumber \\
     & = & \frac{(a\phi ^2+b\theta ^2+c\phi \theta )^{1/2}}{[a-c^2/(4b)]^{1/2}}
\exp \left\{i\left[\arctan \left(\frac{b(\theta /\phi ) + c/2}{(ab-c^2/4)^{1/2}}
\right)\right]\right\} \\
     & = & [1+ic/(4ab-c^2)^{1/2}]\phi + ib\theta /(ab-c^2/4)^{1/2}. \nonumber
\end{eqnarray}

\noindent Thus, the relationships between the coefficients for steady-state
Schr\"{o}dinger wave functions and for the conjugate momentum are given by 
$\alpha = [1+ic/(4ab-c^2)^{1/2}]$  and  $\beta = ib/(ab-c^2/4)^{1/2}$. 
Specifying the initial conditions for the Schr\"{o}dinger wave function along
with the normalization of the Wronskian does indeed determine the coefficients
$(a,b,c)$ for the generalized Hamilton-Jacobi representation.  Hence, the initial
conditions for the wave function determine a unique trajectory for $W'(x)$ in
phase space.  Also, these initial conditions establish to within an integration
constant, K, a unique W which is the generator of motion for a unique trajectory
as a function of time, t, in configuration space as the equation of motion is the
Hamilton-Jacobi transformation equation (often called Jacobi's theorem) $t-\tau
= \partial W/\partial E$ where $\tau$ specifies the epoch.    

We now consider the boundary value problem.  The boundary value problem is not
as simple.  The solutions for boundary value problem, if they exist at all, need
not be unique.  As is well known for bound states, solutions for the
Schr\"{o}dinger wave function do exist for the energy eigenvalues.  Not as well
known, solutions for Hamilton's characteristic function for the trajectory
representation of quantum mechanics exist if the action variable is quantized
[4,11].  Specifically, we consider the bound state problem where  $\psi
\rightarrow 0$  as  $x \rightarrow \pm \infty$.  These are the bound state
eigenfunctions which are unique.  While the Schr\"{o}dinger wave function is
unique for bound states, the conjugate momentum is not [11].  In the generalized
Hamilton-Jacobi representation of quantum mechanics, the boundary conditions for
bound motion manifest a phase-space trajectory with turning points at  $x = \pm
\infty$  (exceptions to this include the infinitely deep square well where the
turning points are at the edges of the well for the Schr\"{o}dinger wave function
does not penetrate the classically forbidden domain).  This is accomplished by
$W' \rightarrow 0$ as $x \rightarrow \pm \infty$.  However, the generalized
Hamilton-Jacobi equation for the bound states is a nonlinear differential
equation that has critical (singular) points at the very location where the
boundary values are applied, i.e.,  $x = \pm \infty$.  By Eq.\ (2),  $W'
\rightarrow 0$  as $x \rightarrow \pm \infty$  because at least one of the
independent solutions, $\phi$ or $\theta$, of the Schr\"{o}dinger equation must
be unbound as  $x \rightarrow \pm \infty$.  As the coefficients satisfy  $a,b >
0$  and  $ab > c^2/4$,  the conjugate momentum exhibits a node as $x \rightarrow
\pm \infty$ for all permitted values of $a$, $b$, and $c$ [11].  Hence, the
boundary values, $W'(x=\pm \infty ) = 0$, for Eq.\ (1) permit non-unique 
phase-space trajectories for $W'$ for energy eigenvalues or quantized action
variables.  Likewise, the trajectories in configuration space are not unique for
the energy eigenvalue as the equation of motion, $t-\tau = \partial W/\partial
E$, specifies a trajectory dependent upon the coefficients $a$, $b$ and $c$. 

We now show that these non-unique trajectories in phase space and configuration
space manifest microstates of the Schr\"{o}dinger wave function.  For bound
states in one dimension, the time-independent Schr\"{o}dinger wave function may
be real except for an inconsequential phase factor [14].  Bound states have the
boundary values that $\psi (x=\pm \infty ) = 0$.  Let us choose $\phi$ to be the
bound solution.   Then $\psi = \alpha \phi$.  The Schr\"{o}dinger wave function
can be represented in trigonometric form as [11]

\begin{eqnarray}
\psi & = & \frac{(2m)^{1/4} cos(W/\hbar )}{(W')^{1/2}[a-c^2/(4b)]^{1/2}}
\nonumber \\
& = & \frac{(a\phi ^2+b\theta ^2+c\phi \theta )^{1/2}}{[a-c^2/(4b)]^{1/2}} \cos
\left[\arctan \left(\frac{b(\theta /\phi ) + c/2}{(ab-c^2/4)^{1/2}}\right)\right]
= \phi .
\end{eqnarray}

\noindent Thus,  $\alpha = 1$  and  $\beta = 0$  for all permitted values of the
set $(a,b,c)$.  Each of these non-unique trajectories of energy $E$ manifest a
microstate of the Schr\"{o}dinger wave function for the bound state.  These
microstates of energy $E$ are specified by the set $(a,b,c)$.  

There is another explanation for our findings.  For the unbound case, $\psi$ is
complex while $W$ and $W'$ are real.  This is consistent with $\psi$ being
proportional to $(W')^{-1/2}\exp (iW/\hbar )$, cf. Eq.\ (4), even though a
relationship exist between $W$ and $W'$.  But $\psi$ may be represented as real
in one dimension for bound states, cf. Eq.\ (5).  Thus, there is some freedom in
choosing $W$ and $W'$ which allows an uncountable number of different
trajectories to be consistent with a particular bound-state eigenfunction.   

We now understand in one dimension the reasons why for bound states the
generalized Hamilton-Jacobi representation describes microstates not detected by
the Schr\"{o}dinger representation while for initial condition problems the two
representations are equivalent.  As the generalized Hamilton-Jacobi
representation provides additional information identifying microstates, we must
conclude that the Schr\"{o}dinger wave function is not an exhaustive description
of nonrelativistic quantum mechanics while the generalized Hamilton-Jacobi
representation is so.

As the generalized Hamilton-Jacobi equation renders information regarding
microstates while the Schr\"{o}dinger equation does not, the two equations are
not equivalent.  The generalized Hamilton-Jacobi equation is the more
fundamental. \\[.05in]

\noindent {\bf 3. HIGHER DIMENSIONS} \\

Our results can be extended to higher dimensions if separation of variables is
permitted.  The existence of various bound-state trajectories for the same 
bound-state wave function in a separable coordinate system manifests a counter
example showing that Schr\"{o}dinger wave function is not the exhaustive
description in higher dimensions.  If the variables, including time, cannot be
separated, the development of closed-form solutions is presently severely
encumbered in either representations.  Furthermore, closed form solutions for
either representation are presently not known in general.  Nevertheless, we may
still discuss the more general problem.  For stationary bound states, closed
Dirichlet boundary conditions are sufficient to determine a unique, stable
solution for the Schr\"{o}dinger wave function.  On the other hand, in the
trajectory representation, the generalized Hamilton-Jacobi representation for the
trajectory in three dimensions is given by the pair of equations [15]

\begin{eqnarray*}
\frac{\partial S}{\partial t} & = & -\frac{(\nabla S)^2}{2m}-V+\frac{\hbar
^2}{2m} \frac{\nabla ^2 R}{R}, \\
\frac{\partial R}{\partial t} & = & -\frac{1}{2m} (R\nabla ^2S+2\nabla R\cdot
\nabla S),
\end{eqnarray*}

\noindent where S is Hamilton's principal function and R is the amplitude
function in three dimensions corresponding to the Milne [4] or Pinney [5]
amplitude function in one dimension.  For systems independent of time, Lee has
simplified the above pair of equations to [16]

\[
\nabla ^2R+(2m/\hbar ^2)(E-V)R-\lambda ^2\kappa ^2R^{-3}=0,
\]

\noindent where $\lambda $ is a normalization constant and $\kappa $ is an
expansion coefficient given by

\[
\kappa = \exp \left(-\int ^{\sigma}\nabla \cdot (\nabla W/|\nabla W|) \, d\sigma
\right),
\]

\noindent where $d\sigma $ is an element of arc length, $d\sigma
^2=dx^2+dy^2+dz^2$ in cartesian coordinates, along the gradient of W. (We note
that in general the gradient of W is not necessarily co-axial with the trajectory
[12].)  The action variable for the generalized Hamilton-Jacobi representation
in higher dimensions still has finite quantization.  Therefore, the contribution
of action along the trajectory in the classically forbidden region must be finite
despite its infinite length out to the turning points at infinity.  Accordingly,
the conjugate momentum projected along the trajectory must decrease extremely
rapidly to zero far in the classically forbidden region.  Hence, the turning
point has the characteristic of a node in the theory of nonlinear differential
equations.  This presents evidence that the trajectory representation for a bound
state has a critical point in general at points where the Dirichlet boundary
conditions are applied to the wave function which in turn indicates that an
uncountable number of microstates (trajectories) for a bound states is possible.
\\[.05in]

\noindent {\bf POSTSCRIPT} \\  

A current investigation, not yet reported, examines the Goos-Hanchen effect in
the trajectory representation of quantum mechanics for reflection off a 
semi-infinite rectangular barrier.  The Schr\"{o}dinger wave function, $\psi $,
for sub-barrier energy, represents neither a bound particle, for such a particle
is not confined to a finite region [17], nor an unbound particle, for $\psi $ is
monotonically dampened in a semi-infinite domain.  For trajectories with sub-
barrier energy, such a barrier generates a nodal singularity in the trajectory
and concurrently a zero in $\psi $ at a point infinitely deep in the barrier. 
This single nodal singularity is sufficient, by itself, to induce microstates of
the Schr\"{o}dinger wave function.  Thus, we conclude that if any nodal
singularity exists in the trajectory representation, then microstates exist and
$\psi $ is not an exhaustive description of nonrelativistic quantum phenomenon.
\\[.05in]     

\noindent {\bf REFERENCES} \\

\begin{enumerate} \leftmargin 0in \itemsep -.06in
\item A. Zwaan, thesis, {\it Intensit\"{a}ten im Ca-Funkenspektrum}, Utrecht,
1929.
\item E. C. Kemble, {\it The Fundamental Principles of Quantum Mechanics with
Elementary Applications} (Dover, New York, 1937), p. 95.
\item A. Mesiah, {\it Quantum Mechanics} (North-Holland, Amsterdam, 1961), Vol.
I, p. 232.
\item W. E. Milne, {\it Phys. Rev.} {\bf 35}, 863 (1930).
\item E. Pinney, {\it Proc. Am. Math. Soc.} {\bf 1}, 681 (1950).
\item C. J. Eliezer and A. Gray, {\it SIAM J. Appl. Math.} {\bf 30}, 463 (1976).
\item H. J. Korsch and H. Laurent, {\it J. Phys.} {\bf B 14}, 4213 (1981).
\item J. Killingbeck, {\it J. Phys.} {\bf A 13}, L231 (1980).
\item H. R. Lewis, Jr., {\it Phys. Rev. Lett.} {\bf 18}, 510 (1967).
\item E. R. Floyd, {\it Phys. Rev.} {\bf D 25}, 1547 (1982).
\item E. R. Floyd, {\it Phys. Rev.} {\bf D 34}, 3246 (1986).
\item E. R. Floyd, {\it Phys. Essays} {\bf 7}, 135 (1994).
\item E. R. Floyd, {\it Ann. Fond. Louis de Broglie} {\bf 20}, 263 (1995).
\item L. D. Landau and E. M. Lifshitz, {\it Quantum Mechanics} (Addison-Wesley,
Reading, Massachusetts, 1958) pp. 52 and 160.
\item D. Bohm, {\it Phys. Rev.} {\bf 85}, 166 (1952).
\item R. A. Lee, {\it J. Phys.} {\bf A 15}, 2761 (1982); {\it Opt. Acta} {\bf
31}, 1119 (1984). 
\item {\it McGraw-Hill Dictionary of Physics and Mathematics}, D. N. Lapedes, ed.
(McGraw-Hill, New York, 1978), p. 108.
\end{enumerate}
   
\end{document}